%% file: apt_jpn_main.tex
\documentclass[12pt,a4paper,notitlepage,fleqn]{article}

\usepackage{setspace}
\usepackage[margin=1in]{geometry}
\usepackage[symbol]{footmisc}

\usepackage[T1]{fontenc}
\usepackage[utf8]{inputenc}
\usepackage{authblk}

\makeatletter

\newcommand{\fboxsubsec}[1]{
	\begin{flushleft}
		#1
	\end{flushleft}
	}
\renewcommand{\subsection}{\@startsection{subsection}{2}{0pt}
	{1ex}
	{0.5ex}
	{\reset@font\it\fboxsubsec}
	}
\makeatother

\title{On the Time-Varying Structure of the Arbitrage Pricing Theory using the Japanese Sector Indices}%

\author{Koichiro Moriya$^{a}$\thanks{\scriptsize Corresponding Author. E-mail: moriya.koichiro@keio.jp, Tel. +81-3-5427-1517.} \ and \ Akihiko Noda$^{b,c}$

{\scriptsize ${}^{a}$ \it Graduate School of Economics, Keio University, 2-15-45 Mita, Minato-ku, Tokyo 108-8345, Japan} 

{\scriptsize ${}^{b}$ \it School of Commerce, Meiji University, 1-1 Kanda-Surugadai, Chiyoda-ku, Tokyo 101-8301, Japan}

{\scriptsize ${}^{c}$ \it Keio Economic Observatory, Keio University, 2-15-45 Mita, Minato-ku, Tokyo 108-8345, Japan}}

\date{This Version: \today}

\renewcommand\thefootnote{\arabic{footnote}}

\usepackage{graphicx}

\usepackage[sort]{natbib}%
\usepackage{amsmath,amssymb}%
\usepackage{ascmac}%
\usepackage{multirow}%
\usepackage{pdflscape}%
\usepackage{subfigmat}

\usepackage{pifont}%
\usepackage{arydshln}%
\usepackage[format=hang]{caption}
\usepackage[all]{xy}

\PassOptionsToPackage{hyphens}{url}
\usepackage{color}
\usepackage[bookmarkstype=toc,
	    colorlinks=true, 
	    pdfborder={0 0 1}, 
	    bookmarks=true, 
	    bookmarksnumbered=true, 
	    citecolor=blue]{hyperref}

\bibpunct{(}{)}{;}{a}{}{,}

\def\hsymbu#1{\smash{\lower1.7ex\hbox{\huge$#1$}}}

\def\ve #1{{\mbox{\boldmath $#1$}}}

\newcommand{\citetapos}[1]{\citeauthor{#1}'s \citeyearpar{#1}}
\newcommand{\citeapos}[2]{\citeauthor{#1}'s (\citeyear{#2})}
\newcommand{\ex}{{\mathbb{E}}}

\def\cov{\mathbb{C}{\rm{ov}}}

\def\ve #1{{\mbox{\boldmath $#1$}}}

\begin{document}

\begin{titlepage}

\renewcommand{\thepage}{}
\renewcommand{\thefootnote}{\fnsymbol{footnote}}

\maketitle

\vspace{-10mm}

\noindent
\hrulefill

\noindent
{\bfseries Abstract:} This paper is the first study to examine the time instability of the APT in the Japanese stock market. In particular, we measure how changes in each risk factor affect the stock risk premiums to investigate the validity of the APT over time, applying the rolling window method to \citetapos{fama1973rre} two-step regression and \citetapos{kamstra2023tra} generalized GRS test. We summarize our empirical results as follows: (1) the changes in monetary policy by major central banks greatly affect the validity of the APT in Japan, and (2) the time-varying estimates of the risk premiums for each factor are also unstable over time, and they are affected by the business cycle and economic crises. Therefore, we conclude that the validity of the APT as an appropriate model to explain the Japanese sector index is not stable over time.\\

\noindent
{\bfseries Keywords:} Time-Varying Coefficient; Arbitrage Pricing Theory; Monetary Policy; Macroeconomic Risk; Generalized GRS Test\\

\noindent
{\bfseries JEL Classification Numbers:} G12; G14; E52.

\noindent
\hrulefill

\end{titlepage}

\bibliographystyle{asa}

\pagebreak

\onehalfspacing

\input{apt_jpn_intro}

\input{apt_jpn_model}

\input{apt_jpn_data}

\input{apt_jpn_empirical}

\input{apt_jpn_conclusion}

\input{apt_jpn_ack}

\clearpage

\input{apt_jpn_main.bbl}
\clearpage

\input{apt_jpn_table_fig}

\end{document}

%% file: apt_jpn_intro.tex
\section{Introduction}\label{sec:apt_intro}

Economists have been investigating which asset pricing model is appropriate for explaining the behavior of the stock returns since \citet{sharpe1964cap} and \citet{lintner1965vra} proposed the capital asset pricing model (CAPM). As a result, there are a large number of models to explain the stock returns. Among them, the intertemporal CAPM of \citet{merton1973aic}, the arbitrage pricing theory (APT) of \citet{ross1976atc}, and \citeapos{fama1993crf}{fama1993crf,fama2015ffa,fama2016daf} multi-factor models (FF models) are often investigated in previous studies. We can interpret the FF models as models that aim to capture the market anomalies such as the size-risk factor and the value-risk factor. However, some previous studies show that the market efficiency or the predictability of stock returns change over time, and they are affected by the various exogenous shocks (see \citet{kim2011srp}, \citet{ito2016eme}, and \citet{noda2016amh}). In other words, the FF models may not sufficiently explain the stock returns  because they do not capture the exogenous shocks. Then we apply the APT to explain the stock returns because the APT can explicitly consider the exogenous shocks and capture the impact of them on the stock returns.

Many previous studies examine the validity of the APT in the U.S. stock market since \citet{chen1986efs}. On the other hand, there are not so many previous studies that test the APT in the Japanese stock market. \citet{hamao1988eea} is the first study to test the APT as an appropriate pricing model in the Japanese stock market from 1975 to 1984 using monthly data. He concludes that the APT is valid in explaining the stock returns in Japan. 
\citet{azeez2006mfe} explore the time instability of the APT by splitting the sample periods into the following three periods: the pre-bubble, bubble, and post-bubble periods, focusing on the risk premium of real estate in the late 1980s using monthly data from January 1989 to December 1998. They show that the APT is supported regardless of the sample period, but real estate is not useful in explaining the returns on the Japanese stock market in any period. \citet{tsuji2007wmi} examines whether risk premiums of other macroeconomic factors (e.g., money supply and gold), which are not considered in previous studies, are incorporated into the returns of Japanese sector indices using monthly data from February 1986 to December 2003. His empirical results show that the risk factors used in previous studies, except for crude oil and exchange rate related-factors, can explain the returns on the Japanese stock market. In addition, they find that the money supply, the gold prices, and the foreign exchange reserves help explain the returns on the Japanese stock market. On the other hand, they conclude that the other factors do not contribute much to explaining Japanese stock market returns.

In addition, we also introduce some related studies that focus on the impact of macroeconomic factors on the Japanese stock market. \citet{kaneko1995rie} examine the effectiveness of economic state variables, used in \citet{chen1986efs} and \citet{hamao1988eea}, in the U.S. and Japanese stock markets based on the vector autoregressive model during the period from January 1975 to December 1993. Their empirical results are not consistent with \citet{hamao1988eea}, but are consistent with \citet{chen1986efs}. \citet{he1998fee} examine whether the stock returns of Japanese multinationals are affected by changes in the exchange rate using monthly data from January 1979 to December 1993. They find that the depreciation (appreciation) of the yen has a positive (negative) effect on the stock returns of Japanese multinationals. \citet{doukas1999pcr} show that the risk premium on the foreign exchange is a significant component for explaining the Japanese sector indices using monthly data from January 1975 to December 1995. Moreover, they find that the risk premium changes over time in response to market conditions. \citet{homma2005ers} examine whether investors carefully monitor the export intensity and net external position of Japanese firms and whether they are adequately reflected in stock prices based on a multifactor model that includes macroeconomic factors using daily data from January 5, 1983 to March 29, 1996. They find that stock investors carefully evaluate the export intensity and net external position of Japanese firms. \citet{thorbecke2020hcc} estimates the sensitivity of risk factors in a multifactor model using daily returns on Japanese sector indices from January 6, 2020, to May 29, 2020. He finds that there is an asymmetry in sensitivity across sectors since the World Health Organization (WHO) declared the COVID-19 global pandemic.

However, there exist two empirical problems with the previous studies for the Japanese stock market. First, the previous studies use monthly data, except for \citet{homma2005ers} and \citet{thorbecke2020hcc}. We believe that the stock returns are likely to be affected by the production activities of domestic firms and monetary policy. Therefore, the industrial production index and the money supply are often used as proxy variables but these variables do not have more high frequent data than monthly, so we cannot ensure a sufficient sample size. Second, the sample periods used in the previous study are too restrictive to examine the long-term validity of the APT. In particular, the sample periods of the sub-samples used in \citet{azeez2006mfe} is less than 100 months, which makes the results skeptical. In addition, \citet{thorbecke2020hcc} examines the impact of the COVID-19 global pandemic on Japanese industries, but the sample period, from January 1 to May 29, 2020, is quite restrictive. Thus, it is difficult to say that we have elucidated the long-term impact of the COVID-19 global pandemic on the Japanese stock market.

Moreover, the previous studies argue that the APT is supported in the Japanese stock market, but the significance and sign condition of the estimates of risk premiums are quite different. We believe that the reason for this inconsistency in the empirical results of the previous studies is that they do not consider the time-varying structure of the APT. In other words, they implicitly assume that the stock market is stable over time. However, some previous studies, such as \citet{kim2011srp}, \citet{ito2016eme}, and \citet{noda2016amh}, find that market efficiency (or stock predictability) varies over time because the environment surrounding the stock market keeps changing due to exogenous shocks such as economic crises. Therefore, we need to consider the time-varying structure of the APT. We then examine the time instability of the validity of the APT over time in the Japanese stock markets. In practice, we apply the rolling window method to \citetapos{fama1973rre} two-step regression using long-term daily data to verify the implicit assumption in previous studies that the market structure is stable over time. Finally, we perform the generalized \citetapos{gibbons1989teg} test developed by \citet{kamstra2023tra} with the rolling window method to examine the validity of the APT over time.

The rest of this paper is organized as follows. Section \ref{sec:apt_model} presents our methodology. Section \ref{sec:apt_data} describes the data used in this paper. Section \ref{sec:apt_empirical} shows our empirical results. Section \ref{sec:apt_conclusion} concludes this study.

%% file: apt_jpn_model.tex
\section{Methodology}\label{sec:apt_model}
This section presents an empirical framework to examine the time instability of \citetapos{ross1976atc} APT using the Japanese sector indices.

\subsection{Preliminaries}\label{subsec:apt_theory}
In the literature of modern finance, it is well known that the CAPM proposed by \citet{sharpe1964cap} and \citet{lintner1965vra} is the most basic framework to explain stock returns. However, the theoretical assumptions to hold the CAPM are quite strong and unrealistic, as shown in \citet{ross1976atc}. In addition, \citet{roll1977cap} also criticizes that in most empirical studies on the CAPM, the market portfolio is limited to stocks (e.g., the Standard \& Poor's 500 and the Tokyo Stock Price Index) and does not include other financial assets.

\citet{ross1976atc} proposes the APT to address the theoretical and empirical problems of the CAPM discussed above. We can interpret the ATP as a theoretically relaxed model that explicitly considers how systematic risks such as business cycle, exchange rates, and policy changes, which cannot be eliminated under the CAPM, affect the returns on financial assets. The APT assumes that each investor (homogeneously) believes that the excess returns on the sector indices follow the $m$-factor generating model below:
\begin{equation}
 r_{i,t} -r_{f,t}= \alpha_{i} + \sum_{j=1}^m \beta_{ij} f_{j,t} +\varepsilon_{i,t}, \ \ i=1,\ldots,n,\ \  j=1,\ldots,m,\ \ m<n,\ \ t=1,\ldots,T,\label{eq1}
\end{equation}
where $r_{i,t}$ is the stock returns for each sector at time $t$, $r_{f,t}$ is the risk-free rate at time $t$, $\alpha_i$ is the constant term for each sector, $\beta_{ij}$ is the sensitivity to the $j$-th factor for each sector, $f_{j,t}$ is the $j$-th factor, which is a systematic risk affecting the risk premium for all sector indices at time $t$, and $\varepsilon_{i,t}$ is the error term specific to $i$-th sector at time $t$. We also make the following assumptions:
\[
 \ex[f_{j,t}]=0, \ \ \ex[\varepsilon_{i,t}]=0, \ \ \cov(\varepsilon_{i,t},\varepsilon_{j,t})=0, \ \ \forall i\neq j.
\]
Suppose that we now invest $x_i$ in any $i$-th sector. We can get $x_i$ for free if short selling is allowed. Specifically, we can construct a portfolio at no cost by investing all the funds if we borrow shares from a brokerage firm to short and sell them.\footnote{Note that we do not consider the existence of the transaction costs here.} The above constraints can be expressed as follows:
\begin{equation}
 \sum_{i=1}^n x_i=0\label{eq2}
\end{equation}
Then we construct a portfolio with no systematic risk $f_j$, i.e., it satisfies the condition in Equation (\ref{eq1}). Since the sensitivity of the systematic risk $\beta_{ij}$ is assumed to be non-zero, the following orthogonal condition is satisfied for $x_i$.
\begin{equation}
 \sum_{i=1}^nx_i\beta_{ij}=0\label{eq3}
\end{equation}
Assuming that we can diversify away the unsystematic risk, any portfolio can be written as follows:
\begin{eqnarray}
 \sum_{i=1}^nx_i(r_{i,t}-r_{f,t})&=&\sum_{i=1}^nx_i\alpha_i+\sum_{i=1}^n\sum_{j=1}^mx_i\beta_{ij}f_{j,t}+\sum_{i=1}^nx_i\varepsilon_{i,t}\nonumber\\
&\simeq&\sum_{i=1}^nx_i\alpha_i+\sum_{i=1}^n\sum_{j=1}^mx_i\beta_{ij}f_{j,t}\nonumber\\
&=&\sum_{i=1}^nx_i\alpha_i\label{eq4}
\end{eqnarray}
Therefore, depending on the combination of $x_i$, it is possible to choose a portfolio with neither systematic nor unsystematic risk. We can understand that all portfolios satisfying the above conditions have no excess returns on average if there are no arbitrage opportunities. That is, if Equations (\ref{eq2}) and (\ref{eq3}) are satisfied, then the following equation is also necessarily satisfied.
\begin{equation}
 \sum_{i=1}^nx_i(r_{i,t}-r_{f,t})=0\Leftrightarrow\sum_{i=1}^nx_i\ex[r_{i,t}]-\sum_{i=1}^nx_ir_{f,t}=0\Leftrightarrow\sum_{i=1}^nx_i\ex[r_{i,t}]=0,\label{eq5}
\end{equation}
where $\ex[r_{i,t}]$ can be expressed as a linear combination of 1 and $\beta_{ij}$. In other words, there exist $m+1$ weights, $\lambda_0,\lambda_1,\ldots,\lambda_m$ such that
\begin{equation}
\ex[r_{i,t}]=\lambda_0+\sum_{j=1}^m \beta_{ij}\lambda_j, \ \ i=1,...,n.\label{eq6}
\end{equation}
Now, we construct a portfolio consisting only of a risk-free asset, we obtain $r_f=\lambda_0$ ($\beta_{i,j}=0$). Hence Equation (\ref{eq6}) can be rewritten as follows:
\begin{equation}
 \ex[r_{i,t}]-r_f=\sum_{j=1}^m \beta_{ij}\lambda_j, \ \ i=1,...,n,\label{eq7}
\end{equation}
The left-hand side of Equation (\ref{eq7}) is the risk premium for the $i$-th sector. If the left-hand side is a risk premium, then the right-hand side must also be an expression of the risk premium. Therefore, $\lambda_j$ can be interpreted as the risk premium for the factor $j \ (j=1,...,m)$. 

\subsection{\citetapos{fama1973rre} Two-Step Regression with Rolling Windows}\label{subsec:apt_empirical}
In this study, we employ \citetapos{fama1973rre} two-step regression with rolling windows as an empirical framework to examine the instability of the APT over time. In particular, we first regress risk premiums against systematic risks to estimate the sensitivity of each risk premium based on Equation (\ref{eq1}). Then we also regress the mean of the risk premium against the estimated sensitivity to estimate the risk premium for each factor based on Equation (\ref{eq7}). In the two-step regression, we employ the feasible generalized least squares (FGLS) estimator instead of the ordinary least squares (OLS) estimator used in many previous studies. This is because the estimation errors are said to be significant when we use the OLS estimator as shown in \citet{shanken1992ebp}.\footnote{\citet{shanken1992ebp} suggests that the GLS or \citetapos{hansen1982lsp} generalized method of moments (GMM) estimator should be used instead of the OLS estimator. However, the GMM estimator have poor finite sample properties as shown in \citet{hansen1996fsp}. Therefore, we employ the FGLS estimator for the two-step regression.}

Equation (\ref{eq1}) can be written in the following matrix form:
\begin{equation}
  \ve{R}_i^{e}=\alpha_i+\ve{F}\ve{\beta}_i+\ve{\varepsilon}_i,\label{eq8}
\end{equation}
where
{\small\[
	\ve{R}_i^{e}=\left(\begin{array}{c}
		 r_{i,1} -r_{f,1} \\
		 r_{i,2} -r_{f,2} \\
		 \vdots\\
		 r_{i,T} -r_{f,T}\\
		       \end{array}\right), \ 
	\ve{F}=\left(\begin{array}{cccc}
		 f_{1,1} &f_{2,1}&\cdots&f_{m,1}\\
		 f_{1,2} &f_{2,2}&\cdots&f_{m,2} \\
		 \vdots  & \vdots&\ddots&\vdots\\
		 f_{1,T} &f_{2,T}&\cdots&f_{m,T} \\
		       \end{array}\right),\ve{\beta}_i=\left(\begin{array}{cccc}
		 \beta_{i1}\\
		 \beta_{i2}\\
		 \vdots\\
		 \beta_{im}\\
		       \end{array}\right), \ 
	\ve{\varepsilon}_i=\left(\begin{array}{c}
		 \varepsilon_{i,1} \\
		 \varepsilon_{i,2} \\
		 \vdots\\
		 \varepsilon_{i,T}\\
		       \end{array}\right),
       \]}
\noindent and $\alpha_i$ and $\ve{\beta}_i$ correspond to the constant terms and the sensitivities in Equation (\ref{eq1}). In addition, we use \citetapos{kamstra2023tra} generalized GRS test to examine whether the APT is a valid model that can explain the risk premiums for Japanese sector indices.\footnote{\citet{kamstra2023tra} point out an over-rejection problem for the test of \citet{gibbons1989teg} and propose the generalized GRS test to solve the problem.} In the generalized GRS test, the null hypothesis is as follows:
\[
 H_0: \ve{\alpha}=\ve{0},\ \ H_1: {\rm{not}} \ H_0
\]
where $\ve{\alpha}$ is $\ve{\alpha}= (\alpha_1, \alpha_2, \ldots, \alpha_n)^\prime$. The generalized GRS test statistics is defined as
\begin{equation}
 \frac{T(T-n-m)}{n(T-m-1)}(1+\ve{\bar{F}}^\prime\ve{\hat{\Omega}}^{-1}\ve{\bar{F}})^{-1}\ve{\hat{\alpha}}^\prime\ve{\hat{\Sigma}}^{-1}\ve{\hat{\alpha}}\sim F_{n,T-n-m},\label{eq9}
\end{equation}
where the test statistics follows the $F$-distribution with degrees of freedom $n$ and $T-n-m$ under the null hypothesis, and each variable is also defined as follows:
\[
\ve{\tilde{F}}_t=\left(\begin{array}{c}
		f_{1,t}\\
		f_{2,t}\\
		\vdots\\
		f_{m,t}
		     \end{array}
\right), \ \ 
\ve{\bar{F}}=\frac{1}{T}\sum_{t=1}^T\ve{\tilde{F}}_t,\ \
\ve{\hat{\Omega}}=\frac{1}{T}\sum_{t=1}^T(\ve{\tilde{F}}_t-\ve{\bar{F}})(\ve{\tilde{F}}_t-\ve{\bar{F}})^\prime, 
\]
\[
\ve{\hat{\alpha}}=\left(\begin{array}{c}
		 \hat{\alpha}_{1} \\
		 \hat{\alpha}_{2} \\
		 \vdots\\
		 \hat{\alpha}_{n}\\
		       \end{array}\right),\ \
 \ve{\hat{\varepsilon}}_t=\left(\begin{array}{c}
		 \hat{\varepsilon}_{1,t} \\
		 \hat{\varepsilon}_{2,t} \\
		 \vdots\\
		 \hat{\varepsilon}_{n,t}\\
		       \end{array}\right),\ \ 
\ve{\hat{\Sigma}}=\frac{1}{T-m-1}\sum_{t=1}^T\ve{\hat{\varepsilon}}_t\ve{\hat{\varepsilon}}_t^\prime,
\]
By estimating Equation (\ref{eq8}), we can obtain $\ve{\hat{\beta}}_i$ and calculate the risk premium $\ve{\lambda}$ for each factor using the obtained $\ve{\hat{\beta}}_i$. In the matrix form, Equation (\ref{eq7}) can be written as
{\small\begin{equation}
    \ex[\ve{R}^e]=\ve{\hat{\beta}}^\prime\ve{\lambda}
    \Leftrightarrow\left(\begin{array}{c}
	\ex[r_{1,t}]-\ex[r_{f,t}] \\
	\ex[r_{2,t}]-\ex[r_{f,t}] \\
	\vdots\\
	\ex[r_{n,t}]-\ex[r_{f,t}] \\
       \end{array}\right)=\left(
	\begin{array}{cccc}
	  \hat{\beta}_{11}&\hat{\beta}_{12}&\cdots&\hat{\beta}_{1m}\\
	  \hat{\beta}_{21}&\hat{\beta}_{22}&\cdots&\hat{\beta}_{2m}\\
	  \vdots&\vdots&\ddots&\vdots\\
	  \hat{\beta}_{n1}&\hat{\beta}_{n2}&\cdots&\hat{\beta}_{nm}\\ 
      \end{array}\right)
      \left(
	\begin{array}{c}
	  \lambda_1\\
	  \lambda_2\\
	  \vdots\\
	  \lambda_m\\
      \end{array}\right)\label{eq10}
\end{equation}}
where
\[
 \ex[\ve{R}^e]=\left(\begin{array}{c}
                \ex[\ve{R}^e_1]\\
		\ex[\ve{R}^e_2]\\
		\vdots\\
		\ex[\ve{R}^e_n]\\
	       \end{array}\right), \ \
	      \ve{\beta}=\left(
		\begin{array}{cccc}
		  \ve{\beta}_1&\ve{\beta}_2&\cdots&\ve{\beta}_n
		\end{array}
	      \right), \ \
	      \ve{\lambda}=\left(\begin{array}{c}
		  \lambda_1\\
		  \lambda_2\\
		  \vdots\\
		  \lambda_m
	      \end{array}\right).\nonumber
\]
Then we can estimate the risk premium $\lambda_j$ for each factor $j$.

Finally, we explain the procedure for testing the time instability of the APT. In modern time series analysis, the rolling-widow method or the state space model is often used to test the time instability of the model. However, it is well known that we have several empirical problems when we estimate the state space model using the maximum likelihood (ML) estimator. A representative one is the pile-up problem which occurs when the ML estimate of the variance of the state equation error is zero, even though its true value is small but not zero, as shown in \citet{sargan1983mle}. Thus, we adopt the rolling-window method to study the time instability of the APT. Suppose that we have time series data with sample size $T$. In the rolling-window method, we set the window width with the sample size of $m<T$ and generate the $T-m+1$ subsample windows by rolling the window each time. We can study the time instability of the APT by using the rolling window method described above. Note that we define the top of the window width as the time at which the estimates and statistics are obtained.

%% file: apt_jpn_data.tex
\section{Data}\label{sec:apt_data}
In this paper, we investigate the time instability of the APT in the Japanese stock market. We first use the daily TOPIX sector indices for 33 industries as portfolios from January 30, 1998 to September 29, 2023. 

Next, we explain how to construct risk factors following the previous studies that examine whether the APT is supported in the Japanese stock market and some related studies that focus on the impact of macroeconomic factors on the Japanese stock market, such as \citet{hamao1988eea}, \citet{kaneko1995rie}, \citet{he1998fee}, \citet{doukas1999pcr}, \citet{homma2005ers}, \citet{azeez2006mfe}, \citet{tsuji2007wmi}, and \citet{thorbecke2020hcc}. We classify risk factors into the following four types: (1) related to stock market factors, (2) commodity market factors, (3) bond market factors, and (4) macroeconomic factors. First, we consider stock market risk factors not only for the domestic stock market but also for the international stock markets, e.g., developed and developing countries. Second, we utilize the forward spread of the commodity to capture the risk in the commodity market because commodities can be alternative financial assets to stocks.\footnote{Previous studies, such as \citet{chen1986efs},  \citet{hamao1988eea}, \citet{kaneko1995rie}, and \citet{tsuji2007wmi}, consider crude oil as alternative financial asset to stock. However, commodities other than crude oil can also be alternative financial assets to stocks as shown in \citet{silvennoinen2013fcc}, we use a composite price index that accounts for all commodity prices.} Third, the yield (government bond) and credit (corporate bond) spreads are used as the bond market risk factor. Fourth, we use the forward premium in the foreign exchange market, unexpected inflation, and the implied volatility index to capture other macroeconomic risk factors.\footnote{Note that our dataset does not include industrial production indices, money supply, and real estate prices among the variables used in previous studies. This is because these variables do not have more high-frequent data than monthly.} 
\begin{center}
<Table \ref{table1} around here>
\end{center}
\noindent Table \ref{table1} summarizes the definitions of the variables used in this paper and how they are calculated. Here, we use the yields of bond trade with repurchase agreement as risk-free rates to calculate the risk premiums following \citet{hamao1988eea}, \citet{kaneko1995rie}, and \citet{tsuji2007wmi}.\footnote{All data except for the returns on the government bond are acquired from the Refinitiv Datastream. Note that the returns on the government bond are obtained from \href{https://www.mof.go.jp/english/policy/jgbs/reference/interest_rate/index.htm}{the website of the Ministry of Finance, Japan}.} We employ the augmented Dickey-Fuller test to check whether each variable satisfies the stationary condition. We also apply \citetapos{schwarz1978edm} Bayesian information criterion to select the optimal lag length for the ADF test. The ADF test rejects the null hypothesis that each variable contains a unit root at the 1\% significance level.\footnote{See Table A.1 in \href{https://at-noda.com/appendix/apt_jpn_appendix.pdf}{the online Appendix} for details.}

%% file: apt_jpn_empirical.tex
\section{Empirical Results}\label{sec:apt_empirical}
We apply \citetapos{fama1973rre} two-step regression with rolling windows to examine the time instability of the APT using the Japanese sector indices. At a stage prior to the time-varying estimation, we employ the two-step regression without rolling windows to investigate whether the APT is supported using the full sample. In particular, we compute the generalized GRS test statistics based on Equation (\ref{eq9}) to examine whether all the constant terms in Equation (\ref{eq8}) are zero. The test statistic is 0.5617, and we cannot reject the null hypothesis that the APT is supported for the entire sample period.\footnote{See Table A.2 in \href{https://at-noda.com/appendix/apt_jpn_appendix.pdf}{the online Appendix} for details.}
\begin{center}
(Table \ref{table2} around here)
\end{center}

Table \ref{table2} presents the time-invariant estimates of the risk premiums based on Equation (\ref{eq10}) and the expected risk premiums calculated from descriptive statistics. The estimates whose sign condition is negative (positive) imply that the risk factors can be alternative (complementary) financial assets to stocks. In practice, we can see that the estimates of $\lambda$ for MKT, UTS, RP, and UI are significantly different from zero at the 5\% level. This means that the four risk factors may help to predict the risk premiums for the returns on the Japanese sector indices throughout our sample period. Furthermore, the sign conditions between the above four estimates and their corresponding expected risk premiums are consistent and these values do not deviate significantly in many factors. This suggests that the risk premiums are correctly estimated for the entire sample period. However, the estimates of the risk premiums in this paper and the previous studies, such as \citet{hamao1988eea}, \citet{azeez2006mfe}, and \citet{tsuji2007wmi}, are inconsistent in the sign conditions and the statistical significance, which means that the estimates of the risk premiums are not stable over time.

Thus we investigate the time instability of the APT in the Japanese stock market using the rolling window method. As mentioned in Section \ref{sec:apt_model}, we apply \citetapos{fama1973rre} two-step regression with rolling windows to investigate whether the APT is time instable. In the time-varying (rolling window) regression, we set the length of the window width to 500, which is approximately two years.\footnote{For the time-varying estimates for the risk factors, see Figures A.1 to A.10 in \href{https://at-noda.com/appendix/apt_jpn_appendix.pdf}{the online Appendix} for details. And we consider window widths from 500 to 1000 in increments of 100 and find that the results are not sensitive to the differences in window widths.} Figure \ref{fig1} shows the time-varying generalized GRS test statistics for each period. We can see that the generalized GRS test statistics is not stable over time. This means that the validity of the APT changes over time.
\begin{center}
 (Figure \ref{fig1} around here)
\end{center}
\noindent We find that the APT is not valid at the 5 percent significance level in the following periods: (1) after June 20, 2012, when the Federal Reserve (Fed) announced an extension of the operation twist program, (2) after April 4, 2013, when the Bank of Japan (BOJ) introduced quantitative and qualitative monetary easing (QQE), (3) after December 13, 2017, when the Fed raised the target range for the federal funds rate from 1--1.25 percent to 1.25--1.5 percent, (4) after July 31, 2018, when the BOJ allowed the yield on the long-term Japanese government bonds (JGBs) to become more volatile, and (5) after September 27, 2018, when the Fed re-raised the target range for the federal funds rate from 1.75--2 percent to 2--2.25 percent. Below we explore the reasons for the invalidity of the APT in the above periods.

First, the deterioration in the validity of the APT in period (1) is most likely due to the Fed's extension of the operation twist program. This program is designed to manipulate the yield curve by selling short term treasuries and buying long term treasuries. \citet{finta2020rps} examine the time-varying risk premium spillovers among the U.S., U.K., German, and Japanese stock markets based on the methods of \citet{diebold2012bgr,diebold2014ont} and \citet{greenwood2021mcg} using tick data from January 2008 to December 2016. They find that the announcement of an extension of the operation twist program on June 20, 2012, increased spillover effects among stock markets. Thus, the extension of the operation twist program may have caused arbitrage opportunities in the Japanese stock market and worsened the validity of the APT.

Second, we believe that the deterioration in the validity of the APT in period (2) is caused by the introduction of the BOJ's QQE. On April 4, 2013, the BOJ decided to double the monetary base, the holdings of JGBs, and the purchases of exchange-traded funds (ETFs). \citet{harada2021bep} examine the impact of the BOJ's ETF purchase program on the Japanese stock market using difference-in-difference analysis. They find that the impact of the program on the Japanese stock market is largest at the introduction of the QQE, even though the BOJ increased the purchase amounts of ETFs as the QQE expanded. Here, the introduction of the QQE was unexpected by most financial market participants. As a result, it may have triggered arbitrage opportunities in the Japanese stock market and worsened the validity of the APT.

Third, the deterioration in the validity of the APT in the periods from (3) to (5) seems to be mainly due to the changes in the Fed's monetary policy. In particular, the Fed raised the target range for the federal funds rate on December 16, 2015, and the generalized GRS statistics started to increase on that day. This increase in the statistics continued until the day when the Fed raised the target range for the federal funds rate from 0.75--1 percent to 1--1.25 percent on June 14, 2017. However, the validity of the APT begins to deteriorate significantly and is rejected at the 5 percent significance level after December 13, 2017, when the Fed re-raised the target range for the federal funds rate from 1--1.25 percent to 1.25--1.5 percent. These results are consistent with \citet{yang2019srg}, who examine that the impact of the Fed's monetary tightening on global financial markets using daily data from March 2011 to the end of 2018. They show that the the Fed's monetary tightening makes Japanes stock price volatile. Thus, the Fed's rate hikes may have caused arbitrage opportunity in the Japanese stock market. 

We also find that the validity of the APT is also affected by the BOJ's monetary policy in this period. In practice, the BOJ allowed the long-term JGB yield to more fluctuate on July 31, 2018. The unexpected change in the volatility of the yield on the long-term JGBs can significantly affect the price of other financial assets. For instance, \citet{hui2022nmy} study the movements of bond yields under yield curve control in Japan. They show that the restoring force of the yield dynamics toward its equilibrium level weakened sharply at the end of September 2018. Their result suggests that arbitrage opportunities between financial assets may have emerged around the periods.

Moreover, our empirical results show that the APT does not hold at the 10 percent significance level in the following periods: (6) after the end of February, 2002, when the BOJ enhenced monetary easing by increasing the outright purchase of long-term government bonds, (7) after April, 12, 2002, when the  Financial Services Agency Japan conducted special inspections and accelerated the disposal of non-performing loans in the Japanese financial sector, (8) after May, 2010, when the European Central Bank (ECB) launched the Securities Market Program (SMP) to address the malfunction of certain gvernment bond markets, (9) after January 29, 2016, when the BOJ began QQE with a negative interest rate, and (10) after the end of November, 2018, when the Fed chairman, Jerome Powell, gave a speech that suggested a pause in the federal fund rate hike.

First, the deterioration in the validity of the APT is more significant in the period (9) than in the periods (6) and (7). \citet{fukuda2018ijn} examines the spillover effects of the BOJ's negative interest rate policy on Asian stock markets. He shows that the introduction of the QQE with a negative interest rate raised serious concerns about the profitability of the Japanese financial sectors. As a result, the Japanese financial sectors explored a new profit opportunity outside Japan, and their impact on Asian stock markets was significantly positive. Therefore, we believe that the QQE with a negative interest rate changed the investment patterns of the Japanese financial sector and caused aribitrage opportunity in this period.

Second, the ECB's introduction of SMP may have caused arbitrage opportunity in
the period (8). \citet{finta2020rps} also find that the ECB's SMP significantly increased spillovers to the Japanese stock market. Therefore, arbitrage opportunities may have emerged after the ECB introduced SMP. Furthermore, we believe that the Fed chairman's speech induced arbitrage opportunity in the period (10). \citet{wen2022mpu} examine the heterogeneous and asymmetric effects of monetary policy uncertainty (MPU) on stock returns in G7 and BRICS countries using monthly data from January 2007 to June 2018. They show that the Japanese stock tends to react significantly to MPU shocks. In practice, the MPU index shows high values after December, 2018. Thus, the Fed chairman's speech implying a pause in the federal fund rate hikes may have increased the MPU and caused arbitrage opportunity in the Japanese stock market. 

In addition, the time-varying generalized GRS test statistics remain in the range where the null hypothesis is not rejected in many periods. However, we can see that the test statistics also change significantly over time, i.e., the validity of the APT has been changing relatively.\footnote{We also confirm the robustness of the results of the generalized GRS test using \citetapos{hansen1997ase} distance (the original HJ-distance) and \citetapos{ren2009ifs} bias-corrected HJ-distance. We obtain the same results that the validity of the APT fluctuates over time as with the results of the generalized GRS test. See Section A.4 in \href{https://at-noda.com/appendix/apt_jpn_appendix.pdf}{the online Appendix} for the detailed results of the robustness check.} In particular, there are some periods where the generalized GRS test statistics have high values but the null hypothesis cannot be rejected in the following periods: (i) the period around 2007, when the subprime mortgage problem became apparent, (i\hspace{-1.2pt}i) after January 22, 2015, when the ECB introduced the quantitative easing, (i\hspace{-1.2pt}i\hspace{-1.2pt}i) after December 14, 2016, when the Fed raised the federal funds rate, and (i\hspace{-1.2pt}v) after April 25, 2019, when the BOJ announced the ETF lending facility. Here, we classify events that affect the behavior of the test statistics into three groups: business cycle, financial crisis, and monetary policy. We regard these events as exogenous shocks.

Our empirical results show that the monetary policy of interest rate manipulation significantly affects the validity of the APT, which is consistent with \citet{inoue2022hdu,inoue2022ise}. They examine the spillover effects of the unconventional monetary policies of the BOJ, the Fed, and the ECB on domestic and global financial markets in the different sample periods using the smooth-transition global VAR model. They show that the monetary easing by these central banks increases the Japanese stock price, and the effects on the Japanese stock market are long-lasting. This implies that the monetary policy changes by these central banks create arbitrage opportunities in the Japanese stock market. Moreover, our empirical results is also consistent with \citet{noda2016amh}, who identifies the periods when the Japanese stock market is efficient based on \citeapos{ito2016eme}{ito2016eme,ito2022aae}. He shows that the efficiency of the Japanese stock market changes in response to the business cycle and exogenous shocks such as economic crises and wars.

\begin{center}
 (Figure \ref{fig2} around here)
\end{center}

Figure \ref{fig2} shows the time-varying estimates for the risk premiums. We recognize that the behavior of each risk premium is not stable over time and is significantly affected by the business cycle and exogenous shocks such as financial crises, the COVID-19 global pandemic, and the Russian invasion of Ukrainein in 2022. Here, the risk factor whose sign condition is positive (negative) is defined as financial assets complementary (alternative) to stocks. Therefore, our empirical results imply that the relationship between each risk factor and the Japanese sector indices changes over time. The detailed estimation results are summarized below.

First, the time-varying estimates of $\lambda_{MKT}$ is upward (downward) sloping during expansions (recessions). This result is consistent with \citet{bansal2021tsr}, who examine the term structure of expected equity risk premia over time in the U.S., Europe, and Japan. In addition, our empirical results show that the volatilities of $\lambda_{WORLD}$ and $\lambda_{EMERGE}$ increase in response to financial crises. This result is consistent with \citet{choudhry2007cst}, who examine the changes in the long-run relationships between the stock prices of Far East countries, Japan, and the U.S. around the Asian financial crisis. They suggest that the relationships between the Japanese and other stock markets become closer during financial crises.

Second, we find that the time-varying estimates of $\lambda_{CMD}$ is not statistically significant over the sample periods, but its volatility is significantly affected by exogenous shocks. This result is consistent with \citet{thuraisamy2013rae}, who investigate the volatility interactions between 14 Asian stock markets, including the Japanese stock market, and dominant commodity markets. They show that the volatility dynamics changes substantially during economic crises. Furthermore, \citet{zhang2021com} investigate the return and volatility spillover between the COVID-19 global pandemic in 2020, the crude oil market, and the stock market. They find that the volatility spillovers between the crude oil and stock markets increase significantly more during the COVID-19 global pandemic than during the global financial crisis. Therefore, this insights the volatility of $\lambda_{CMD}$ may increase significantly during the COVID-19 global pandemic.

Third, the time-varying estimates of $\lambda_{RP}$ and $\lambda_{UTS}$ are unstable over time and behave oppositely. The correlation coefficient for these time-varying estimates is $-0.7763$. In addition, the behavior of these two variables is affected by the business cycle. However, there are differences in the extent to which the business cycle affects these variables in that $\lambda_{RP}$ is more affected by the business cycle than $\lambda_{UTS}$. This result is consistent with \citet{okimoto2017tsc,okimoto2022csc}, who examine the predictability of the business cycle in Japan using the term structure of credit spreads and the government bond yield spread. Their empirical results show that the term structure of credit spreads is more helpful in predicting the business cycle in Japan than the government bond yield spread.

Fourth, the time-varying estimates of $\lambda_{UYEN}$ is not statistically significant over the sample periods, which is consistent with \citet{hamao1988eea} and \citet{tsuji2007wmi}. Moreover, our empirical results reveal that the behavior of $\lambda_{UYEN}$ is significantly affected by changes in monetary policy, and $\lambda_{UYEN}$ converges to zero during the COVID-19 global pandemic. This suggests that the UYEN does not contribute to the formation of the risk premium of the Japanese sector indices during the COVID-19 global pandemic. \citet{aslam2020oef} examine the efficiency of foreign exchange markets during the COVID-19 pandemic. Their empirical results show that the Japanese yen against the U.S. dollar was highly efficient before the COVID-19 global pandemic periods, but the efficiency of the exchange rate decreases tremendously during the COVID-19 global pandemic. Therefore, the inefficiency of the foreign exchange market may be the reason why the UYEN does not contribute to the appropriate formation of the risk premium in the Japanese sector indices during the COVID-19 global pandemic. 

Fifth, we find that the time-varying estimates of $\lambda_{UI}$ is not statistically significant and stable in many periods. However, the sign condition of $\lambda_{UI}$ becomes unstable and the volatility of the estimates increases after the COVID-19 global pandemic. We believe that the changes in the BOJ's monetary policy cause the rapid fluctuation of $\lambda_{UI}$. The BOJ enhanced monetary easing on April 27, 2020, in response to the COVID-19 global pandemic. In practice, the BOJ decided to purchase the necessary amount of JGBs, without setting a ceiling, so that the yield on the 10-year JGB would remain around zero percent. In addition, the BOJ declared that on March 19, 2021, the range of fluctuation of the ten-year JGB yield will be between around plus and minus 0.25 percent from the target level. After that, on December 20, 2022, the BOJ announced that the range of fluctuation of the ten-year JGB yield will change from between around plus and minus 0.25 percent to between around plus and minus 0.5 percent. As a result, we believe that it had a significant impact on the estimate of $\lambda_{UI}$ and its volatility.

Last, the time-varying estimates of $\lambda_{VIX}$ is not stable over time, and the significance is greatly affected by exogenous shocks such as the changes in BOJ's monetary policy, financial crises, and wars. This result is consistent with \citet{hong2023tvc}, who examine the time-varying structure of the causal relationship between economic policy uncertainty and the stock market index in the G7 countries and other emerging markets using the time-varying Granger causality tests. Thye show that economic policy uncertainty often has an impact on the stock market during periods of major international crises or national events. Here, they use the economic policy uncertainty (EPU) index constructed by \citet{baker2016meu} as a proxy variable, which is known to be significantly related to the VIX (see \citet{shaikh2019ore} for details). Thus, we can consider that the EPU indices as a proxy for the VIX and our empirical results are plausible.

%% file: apt_jpn_conclusion.tex
\section{Conclusion}\label{sec:apt_conclusion}

This paper is the first study to investigate the time instability of the APT in the Japanese stock market. Previous studies, such as \citet{hamao1988eea}, \citet{azeez2006mfe}, and \citet{tsuji2007wmi}, estimate the time-invariant risk premium under the implicit assumption that the market structure is stable over time. Then we employ \citetapos{fama1973rre} two-step regression with rolling windows and \citetapos{kamstra2023tra} generalized GRS test to examine the validity of the assumption in the previous studies. In addition, we measure how changes in each risk factor affect the risk premiums in the Japanese stock market based on the above time-varying procedure. Our findings are summarized below.

First, we show that the validity of the APT varies over time in the Japanese stock market. In particular, we find that the changes in the monetary policies of the BOJ, ECB, and Fed significantly affect the effectiveness of the APT. Moreover, the business cycle and economic crises have an impact on the magnitude of the validity of the APT. Second, we show that the time-varying estimates of the risk premiums for each factor are also unstable over time, and they are affected by the business cycle and economic crises. This implies that the relationship between stock returns and risk factors varies over time.

Thus, while previous studies assume that the validity of APT is stable over time, our results suggest that the validity varies over time. We conclude that the implicit assumption of previous studies that the market structure is stable over time is not reasonable.

%% file: apt_jpn_ack.tex
\section*{Acknowledgements}
The author would like to thank Kohei Aono, Alok Bhargava, Ryoji Hiraguchi, Daisuke Nagakura, Tatsuyoshi Okimoto, Rui Ota, Shiba Suzuki, Yoichi Tsuchiya, Tatsuma Wada, Tomoaki Yamada and the conference participants at the 98th Annual Conference of the Western Economic Association International and the Japan Society of Monetary Economics 2023 Autumn Meeting for their helpful comments and suggestions. Noda is grateful for the financial assistance provided by the Japan Society for the Promotion of Science Grant in Aid for Scientific Research (grant numbers 19K13747 and 23H00838) and the Japan Science and Technology Agency, Moonshot Research \& Development Program (grant number: JPMJMS2215). All data and programs used are available upon request.

%% file: apt_jpn_table_fig.tex
\setcounter{table}{0}
\renewcommand{\thetable}{\arabic{table}}

\clearpage

\begin{table}[p]
\caption{Variable Definitions}\label{table1}
\begin{center}
 \resizebox{15cm}{!}{\begin{tabular}{l|c|c} \hline\hline
Variables & Details & Data Source\\\hline
\underline{Risk Premiums} &  & \\
\ \ FAF & TOPIX Fishery, Agriculture \& Forestry & Tokyo Stock Exchange\\
\ \ FOD & TOPIX Foods & Tokyo Stock Exchange\\
\ \ MIN & TOPIX Mining & Tokyo Stock Exchange\\
\ \ OIL & TOPIX Oil and Coal Products & Tokyo Stock Exchange\\
\ \ CON & TOPIX Construction & Tokyo Stock Exchange\\
\ \ MET & TOPIX Metal Products & Tokyo Stock Exchange\\
\ \ GLC & TOPIX Glass and Ceramics Products & Tokyo Stock Exchange\\
\ \ TEX & TOPIX Textiles and Apparels & Tokyo Stock Exchange\\
\ \ PUL & TOPIX Pulp and Paper & Tokyo Stock Exchange\\
\ \ CHE & TOPIX Chemicals & Tokyo Stock Exchange\\
\ \ PHA & TOPIX Pharmaceutical & Tokyo Stock Exchange\\
\ \ RUB & TOPIX Rubber Products & Tokyo Stock Exchange\\
\ \ TEQ & TOPIX Transportation Equipment & Tokyo Stock Exchange\\
\ \ IRS & TOPIX Iron and Steel & Tokyo Stock Exchange\\
\ \ NFM & TOPIX Nonferrous Metals & Tokyo Stock Exchange\\
\ \ MAC & TOPIX Machinery & Tokyo Stock Exchange\\
\ \ ELA & TOPIX Electric Appliances & Tokyo Stock Exchange\\
\ \ PRE & TOPIX Precision Instruments & Tokyo Stock Exchange\\
\ \ OTH & TOPIX Other Products & Tokyo Stock Exchange\\
\ \ INF & TOPIX Information \& Communication & Tokyo Stock Exchange\\
\ \ SER & TOPIX Services & Tokyo Stock Exchange\\
\ \ ELP & TOPIX Electric Power and Gas & Tokyo Stock Exchange\\
\ \ LTP & TOPIX Land Transportation & Tokyo Stock Exchange\\
\ \ MTP & TOPIX Marine Transportation & Tokyo Stock Exchange\\
\ \ ATP & TOPIX Air Transportation & Tokyo Stock Exchange\\
\ \ WHS & TOPIX Warehousing and Harbor Transportation & Tokyo Stock Exchange\\
\ \ WHO & TOPIX Wholesale Trade & Tokyo Stock Exchange\\
\ \ RET & TOPIX Retail Trade & Tokyo Stock Exchange\\
\ \ BAK & TOPIX Banks & Tokyo Stock Exchange\\
\ \ SEC & TOPIX Securities and Commodities Futures & Tokyo Stock Exchange\\
\ \ INS & TOPIX Insurance & Tokyo Stock Exchange\\
\ \ OFB & TOPIX Other Financing Business & Tokyo Stock Exchange\\
\ \ RES & TOPIX Real Estate & Tokyo Stock Exchange\\\hline
\underline{Risk Factors} &  & \\
\ \ MKT & Difference between Returns on TOPIX and Gensaki Rate & Tokyo Stock Exchange and Tanshi Kyokai\\
\ \ WORLD & MSCI World Index (Developed 23 countries) & Morgan Stanley\\
\ \ EMERGE & MSCI Emerging Markets Index (Developing 26 countries) & Morgan Stanley\\
\ \ CMD & Forward Spread between Spot and Three-month S\&P GSCI Commodity Index  & Standard \& Poors\\
\ \ UTS & Yield Spread between 10 year Japanese Government Bonds (JGB) and Gensaki Rate & MOF Japan and Tanshi Kyokai\\
\ \ RP & Credit Spread between S\&P 10+years corporate bond and 10 year JGB & Standard \& Poors and MOF Japan\\
\ \ UYEN & Forward Premium between the Spot and Three-month Yen/U.S. Dollar Exchange Rates & Bank of Japan and MUFG Bank, Ltd.\\
\ \ UI & Unexpected Inflation Calculated in Accordance with \citet{chen1986efs} & Tanshi Kyokai and JP Morgan\\
\ \ VIX & Nikkei Stock Average Volatility Index & Nikkei\\\hline\hline
\end{tabular}}
\resizebox{14cm}{!}{\begin{minipage}{650pt}
{\underline{Notes:}}
 \begin{itemize}
  \item[(1)] We calculate UTS, RP, and UYEN following \citet{chen1986efs}.
  \item[(2)] We use \citetapos{fama1984aci} method to obtain the variable of UI.
 \end{itemize}
\end{minipage}}
\end{center}
\end{table}

\clearpage

\begin{table}
\caption{The Estimated and Expected Risk Premiums (Whole Sample)}\label{table2}
\begin{center}
 \begin{tabular}{c|cc}\hline\hline
 & Estimated & Expected \\\hline
\multirow{2}{*}{$\lambda_{MKT}$} & $-0.0007$ & \multirow{2}{*}{$-0.0007$} \\
                     & [0.0000] & \\ 
\multirow{2}{*}{$\lambda_{WORLD}$} & $-0.0003$ & \multirow{2}{*}{$-0.0006$} \\
                     & [0.0014] & \\
\multirow{2}{*}{$\lambda_{EMERGE}$} & 0.0007 & \multirow{2}{*}{$-0.0007$} \\
                     & [0.0009] & \\
\multirow{2}{*}{$\lambda_{CMD}$} & 0.0006 & \multirow{2}{*}{$-0.0008$} \\
                     & [0.0004] & \\
\multirow{2}{*}{$\lambda_{UTS}$} & 0.0025 & \multirow{2}{*}{$0.0074$} \\
                     & [0.0011] & \\
\multirow{2}{*}{$\lambda_{RP}$} & $-0.0028$ & \multirow{2}{*}{$-0.0097$} \\
                     & [0.0012] & \\
\multirow{2}{*}{$\lambda_{UYEN}$} & $0.0012$ & \multirow{2}{*}{$-0.0009$} \\
                     & [0.0009] & \\
\multirow{2}{*}{$\lambda_{UI}$} & $-0.0002$ & \multirow{2}{*}{$-0.0008$} \\
                     & [0.0001] & \\
\multirow{2}{*}{$\lambda_{VIX}$} & $0.0022$ & \multirow{2}{*}{$-0.0009$} \\
                     & [0.0041] & \\\hline\hline
 \end{tabular}
\resizebox{12cm}{!}{\begin{minipage}{450pt}
{\underline{Notes:}}
 \begin{itemize}
  \item[(1)] ``Estimated'' and ``Expected'' denote the $\hat\lambda_i$ and the sample mean $\ex[\lambda_i]$, respectively.
  \item[(2)] The robust standard errors for the GLS estimation are shown in brackets.
  \item[(3)] R version 4.3.1 was used to compute the estimates and statistics.
 \end{itemize}
\end{minipage}}
\end{center}
\end{table}

\clearpage

\begin{figure}[p]
  \caption{Time-Varying Generalized GRS Statistics}\label{fig1}
   \begin{center}
    \includegraphics[scale=0.8]{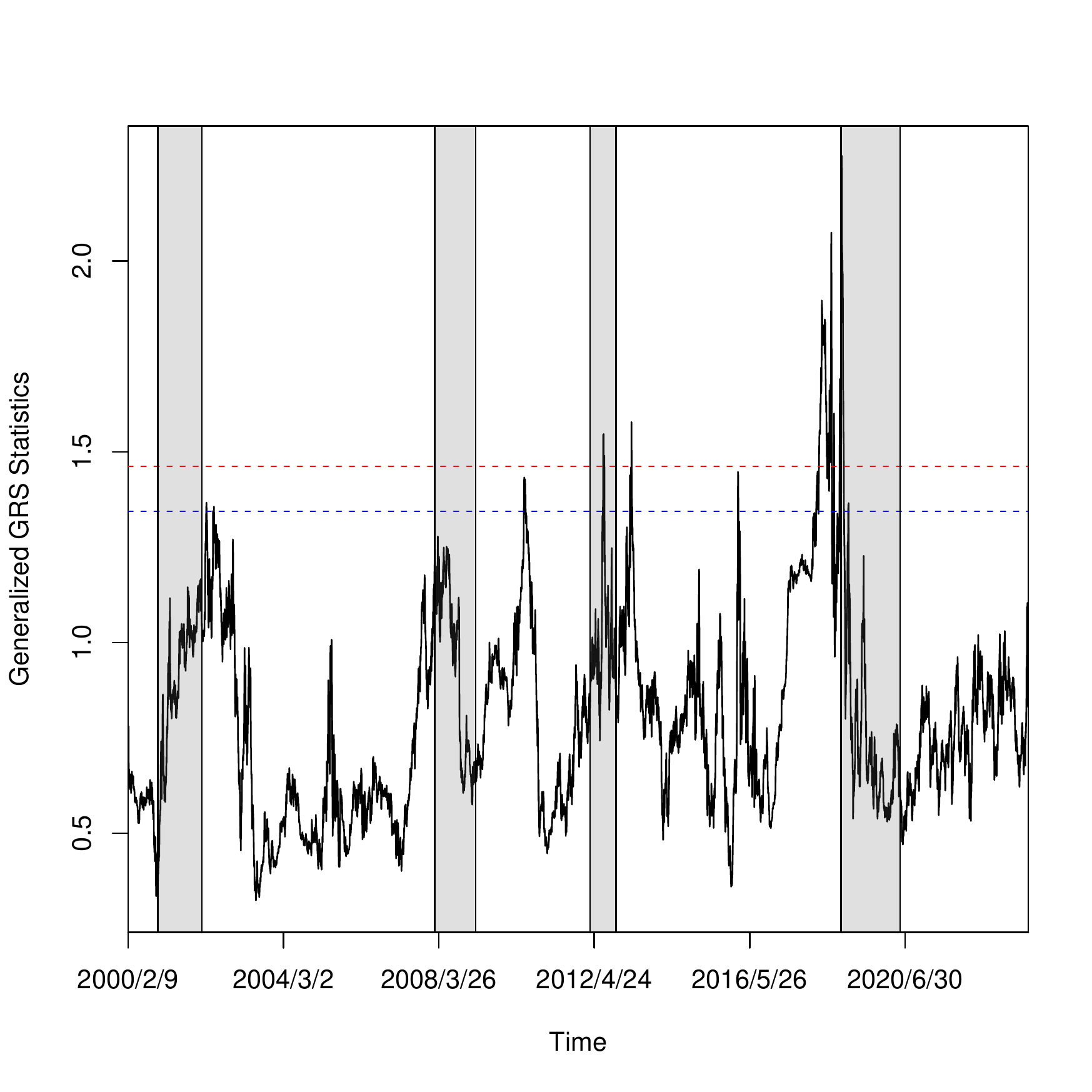}
    \resizebox{15cm}{!}{\begin{minipage}{600pt}
	\underline{Notes:}
	\begin{itemize}
	  \item[(1)] The dashed red and blue lines denote the critical value at the 5\% and 10\% significance levels for the generalized GRS test, respectively.
	  \item[(2)] The shade areas are recessions as defined by the Cabinet Office Japan ``\href{https://www.esri.cao.go.jp/en/stat/di/di-e.html}{Indexes of Business Conditions}.'' 
          \item[(3)] R version 4.3.2 was used to compute the statistics.
	\end{itemize}
    \end{minipage}}
   \end{center}
 \end{figure}

\clearpage

\begin{landscape}
\begin{figure}[p]
\caption{Time-Varying Estimates of Risk Premiums}\label{fig2}
   \begin{center}
    \includegraphics[scale=0.3]{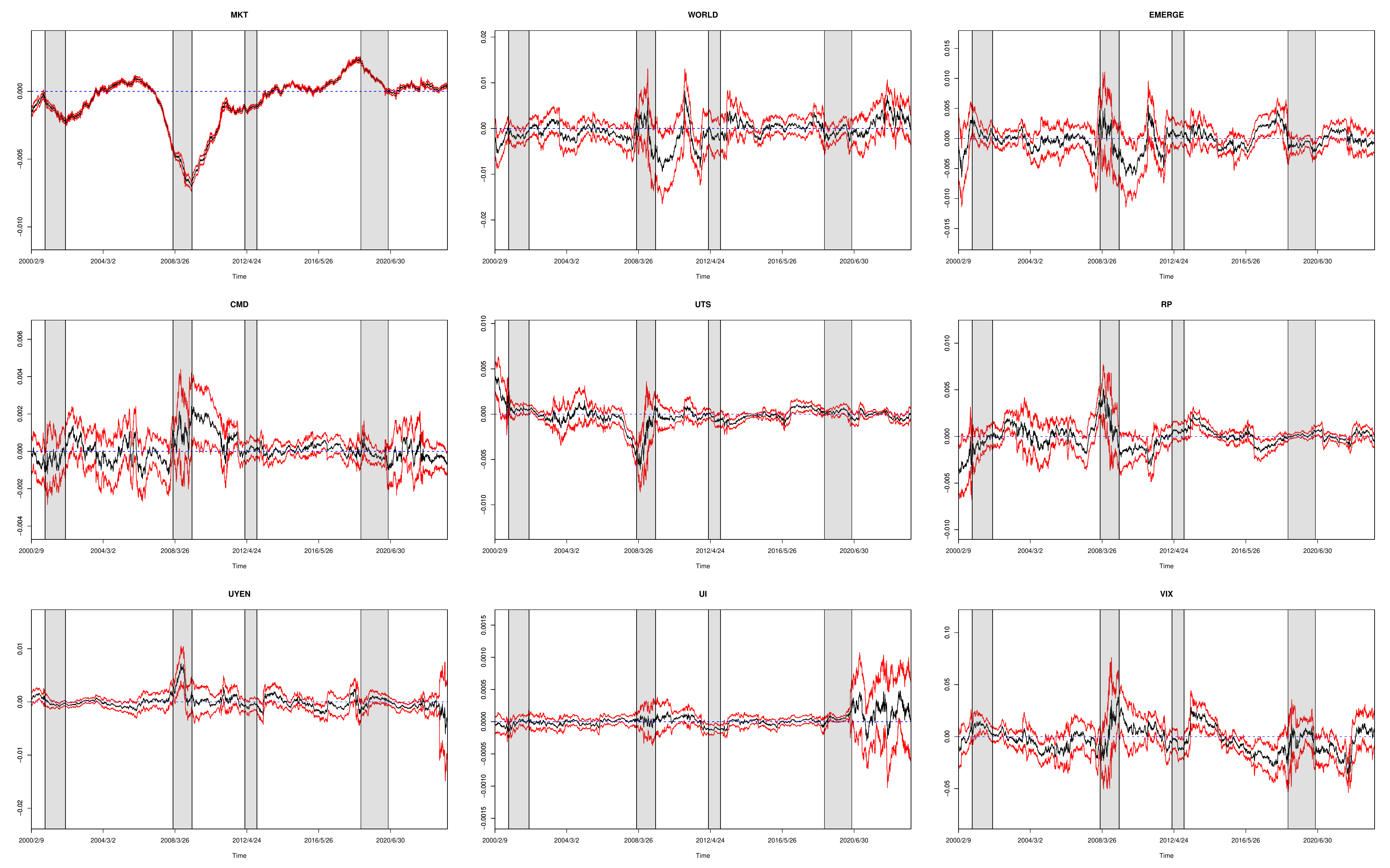}
    \resizebox{18cm}{!}{\begin{minipage}{700pt}
	\underline{Notes:}
	\begin{itemize}
	  \item[(1)] The dashed red lines represent the 95\% confidence intervals of the estimates.
	  \item[(2)] The shade areas are recessions as defined by the Cabinet Office Japan ``\href{https://www.esri.cao.go.jp/en/stat/di/di-e.html}{Indexes of Business Conditions}.'' 
          \item[(3)] R version 4.3.2 was used to compute the estimates.
	\end{itemize}
    \end{minipage}}%
   \end{center}
\end{figure}
\end{landscape}

\clearpage

%% file: apt_jpn_main.bbl
\begin{thebibliography}{54}
\newcommand{\enquote}[1]{``#1''}
\expandafter\ifx\csname natexlab\endcsname\relax\def\natexlab#1{#1}\fi

\bibitem[{Aslam et~al.(2020)Aslam, Aziz, Nguyen, Mughal, and
  Khan}]{aslam2020oef}
Aslam, F., Aziz, S., Nguyen, D.~K., Mughal, K.~S., and Khan, M. (2020),
  \enquote{On the Efficiency of Foreign Exchange Markets in Times of the
  COVID-19 Pandemic,} \textit{Technological Forecasting \& Social Change}, 161,
  120261.

\bibitem[{Azeez and Yonezawa(2006)}]{azeez2006mfe}
Azeez, A.~A. and Yonezawa, Y. (2006), \enquote{Macroeconomic Factors and the
  Empirical Content of the Arbitrage Pricing Theory in the Japanese Stock
  Market,} \textit{Japan and the World Economy}, 18, 568--591.

\bibitem[{Baker et~al.(2016)Baker, Bloom, and Davis}]{baker2016meu}
Baker, S.~R., Bloom, N., and Davis, S.~J. (2016), \enquote{Measuring Economic
  Policy Uncertainty,} \textit{Quarterly Journal of Economics}, 131,
  1593--1636.

\bibitem[{Bansal et~al.(2021)Bansal, Miller, Song, and Yaron}]{bansal2021tsr}
Bansal, R., Miller, S., Song, D., and Yaron, A. (2021), \enquote{The Term
  Structure of Equity Risk Premia,} \textit{Journal of Financial Economics},
  142, 1209--1228.

\bibitem[{Chen et~al.(1986)Chen, Roll, and Ross}]{chen1986efs}
Chen, N., Roll, R., and Ross, S.~A. (1986), \enquote{Economic Forces and the
  Stock Market,} \textit{Journal of Business}, 59, 383--403.

\bibitem[{Choudhry et~al.(2007)Choudhry, Lu, and Peng}]{choudhry2007cst}
Choudhry, T., Lu, L., and Peng, K. (2007), \enquote{Common Stochastic Trends
  among Far East stock Prices: Effects of the Asian Financial Crisis,}
  \textit{International Review of Financial Analysis}, 16, 242--261.

\bibitem[{Diebold and Yilmaz(2012)}]{diebold2012bgr}
Diebold, F.~X. and Yilmaz, K. (2012), \enquote{Better to Give than to Receive:
  Predictive Directional Measurement of Volatility Spillovers,}
  \textit{International Journal of Forecasting}, 28, 57--66.

\bibitem[{Diebold and Yilmaz(2014)}]{diebold2014ont}
--- (2014), \enquote{On the Network Topology of Variance Decompositions:
  Measuring the Connectedness of Financial Firms,} \textit{Journal of
  Econometrics}, 182, 119--134.

\bibitem[{Doukas et~al.(1999)Doukas, Hall, and Lang}]{doukas1999pcr}
Doukas, J., Hall, P.~H., and Lang, L. H.~P. (1999), \enquote{The Pricing of
  Currency Risk in Japan,} \textit{Journal of Banking \& Finance}, 23, 1--20.

\bibitem[{Fama and French(1993)}]{fama1993crf}
Fama, E.~F. and French, K.~R. (1993), \enquote{Common Risk Factors in the
  Returns on Stocks and Bonds,} \textit{Journal of Financial Economics}, 33,
  3--56.

\bibitem[{Fama and French(2015)}]{fama2015ffa}
--- (2015), \enquote{A Five-Factor Asset Pricing Model,} \textit{Journal of
  Financial Economics}, 116, 1--22.

\bibitem[{Fama and French(2016)}]{fama2016daf}
--- (2016), \enquote{Dissecting Anomalies with a Five-Factor Model,}
  \textit{Review of Financial Studies}, 29, 69--103.

\bibitem[{Fama and Gibbons(1984)}]{fama1984aci}
Fama, E.~F. and Gibbons, M.~R. (1984), \enquote{A Comparison of Inflation
  Forecasts,} \textit{Journal of Monetary Economics}, 13, 327--348.

\bibitem[{Fama and MacBeth(1973)}]{fama1973rre}
Fama, E.~F. and MacBeth, J.~D. (1973), \enquote{Risk, Return, and Equilibrium:
  Empirical Tests,} \textit{Journal of Political Economy}, 81, 607--636.

\bibitem[{Finta and Aboura(2020)}]{finta2020rps}
Finta, M.~A. and Aboura, S. (2020), \enquote{Risk Premium Spillovers among
  Stock Markets: Evidence from Higher-Order Moments,} \textit{Journal of
  Financial Markets}, 49, 100533.

\bibitem[{Fukuda(2018)}]{fukuda2018ijn}
Fukuda, S. (2018), \enquote{Impacts of Japan's Negative Interest Rate Policy on
  Asian Financial Markets,} \textit{Pacific Economic Review}, 23, 67--79.

\bibitem[{Gibbons et~al.(1989)Gibbons, Ross, and Shanken}]{gibbons1989teg}
Gibbons, M.~R., Ross, S.~A., and Shanken, J. (1989), \enquote{A Test of the
  Efficiency of a Given Portfolio,} \textit{Econometrica}, 57, 1121--1152.

\bibitem[{Greenwood-Nimmo et~al.(2021)Greenwood-Nimmo, Nguyen, and
  Shin}]{greenwood2021mcg}
Greenwood-Nimmo, M., Nguyen, V.~H., and Shin, Y. (2021), \enquote{Measuring the
  Connectedness of the Global Economy,} \textit{International Journal of
  Forecasting}, 37, 899--919.

\bibitem[{Hamao(1988)}]{hamao1988eea}
Hamao, Y. (1988), \enquote{An Empirical Examination of the Arbitrage Pricing
  Theory: Using Japanese Data,} \textit{Japan and the World Economy}, 1,
  45--61.

\bibitem[{Hansen(1982)}]{hansen1982lsp}
Hansen, L.~P. (1982), \enquote{Large Sample Properties of Generalized Method of
  Moments Estimators,} \textit{Econometrica}, 50, 1029--1054.

\bibitem[{Hansen et~al.(1996)Hansen, Heaton, and Yaron}]{hansen1996fsp}
Hansen, L.~P., Heaton, J., and Yaron, A. (1996), \enquote{Finite-Sample
  Properties of Some Alternative GMM Estimators,} \textit{Journal of Business
  \& Economic Statistics}, 14, 262--280.

\bibitem[{Hansen and Jagannathan(1997)}]{hansen1997ase}
Hansen, L.~P. and Jagannathan, R. (1997), \enquote{Assessing Specification
  Errors in Stochastic Discount Factor Models,} \textit{Journal of Finance},
  52, 557--590.

\bibitem[{Harada and Okimoto(2021)}]{harada2021bep}
Harada, K. and Okimoto, T. (2021), \enquote{The BOJ's ETF Purchases and Its
  Effects on Nikkei 225 Stocks,} \textit{International Review of Financial
  Analysis}, 77, 101826.

\bibitem[{He and Ng(1998)}]{he1998fee}
He, J. and Ng, K. (1998), \enquote{The Foreign Exchange Exposure of Japanese
  Multinational Corporations,} \textit{Journal of Finance}, 53, 733--753.

\bibitem[{Homma et~al.(2005)Homma, Tsutsui, and Benzion}]{homma2005ers}
Homma, T., Tsutsui, Y., and Benzion, U. (2005), \enquote{Exchange Rate and
  Stock Prices in Japan,} \textit{Applied Financial Economics}, 15, 469--478.

\bibitem[{Hong et~al.(2023)Hong, Zhang, and Zhang}]{hong2023tvc}
Hong, Y., Zhang, R., and Zhang, F. (2023), \enquote{Time-Varying Causality
  Impact of Economic Policy Uncertainty on Stock Market Returns: Global
  Evidence from Developed and Emerging Countries,} \textit{International Review
  of Financial Analysis}, 91, 102991.

\bibitem[{Hui et~al.(2022)Hui, Wong, and Lo}]{hui2022nmy}
Hui, C., Wong, A., and Lo, C. (2022), \enquote{A Note on Modelling Yield Curve
  Control: A Target-Zone Approach,} \textit{Finance Research Letters}, 49,
  103076.

\bibitem[{Inoue and Okimoto(2022{\natexlab{a}})}]{inoue2022hdu}
Inoue, T. and Okimoto, T. (2022{\natexlab{a}}), \enquote{How does
  Unconventional Monetary Policy Affect the Global Financial Markets?}
  \textit{Empirical Economics}, 62, 1013--1036.

\bibitem[{Inoue and Okimoto(2022{\natexlab{b}})}]{inoue2022ise}
--- (2022{\natexlab{b}}), \enquote{International Spillover Effects of
  Unconventional Monetary Policies of Major Central Banks,}
  \textit{International Review of Financial Analysis}, 79, 101968.

\bibitem[{Ito et~al.(2016)Ito, Noda, and Wada}]{ito2016eme}
Ito, M., Noda, A., and Wada, T. (2016), \enquote{The Evolution of Stock Market
  Efficiency in the US: A Non-Bayesian Time-Varying Model Approach,}
  \textit{Applied Economics}, 48, 621--635.

\bibitem[{Ito et~al.(2022)Ito, Noda, and Wada}]{ito2022aae}
--- (2022), \enquote{An Alternative Estimation Method for Time-Varying
  Parameter Models,} \textit{Econometrics}, 10, 23.

\bibitem[{Kamstra and Shi(2023)}]{kamstra2023tra}
Kamstra, M, J. and Shi, R. (2023), \enquote{Testing and Ranking of Asset
  Pricing Models Using the GRS Statistic,} Available at {\it{Social Science
  Research Network}}: \url{http://dx.doi.org/10.2139/ssrn.4374415}.

\bibitem[{Kaneko and Lee(1995)}]{kaneko1995rie}
Kaneko, T. and Lee, B. (1995), \enquote{Relative Importance of Economic Factors
  in the U.S. and Japanese Stock Markets,} \textit{Journal of the Japanese and
  International Economies}, 9, 290--307.

\bibitem[{Kim et~al.(2011)Kim, Shamsuddin, and Lim}]{kim2011srp}
Kim, J.~H., Shamsuddin, A., and Lim, K.~P. (2011), \enquote{Stock Return
  Predictability and the Adaptive Markets Hypothesis: Evidence from
  Century-Long U.S. Data,} \textit{Journal of Empirical Finance}, 18, 868--879.

\bibitem[{Lintner(1965)}]{lintner1965vra}
Lintner, J. (1965), \enquote{The Valuation of Risk Assets and the Selection of
  Risky Investments in Stock Portfolios and Capital Budgets,} \textit{Review of
  Economics and Statistics}, 47, 13--37.

\bibitem[{Merton(1973)}]{merton1973aic}
Merton, R.~C. (1973), \enquote{An Intertemporal Capital Asset Pricing Model,}
  \textit{Econometrica}, 41, 867--887.

\bibitem[{Noda(2016)}]{noda2016amh}
Noda, A. (2016), \enquote{A Test of the Adaptive Market Hypothesis using A
  Time-Varying AR Model in Japan,} \textit{Finance Research Letters}, 17,
  66--71.

\bibitem[{Okimoto and Takaoka(2017)}]{okimoto2017tsc}
Okimoto, T. and Takaoka, S. (2017), \enquote{The Term Structure of Credit
  Spreads and Business Cycle in Japan,} \textit{Journal of the Japanese and
  International Economies}, 45, 27--36.

\bibitem[{Okimoto and Takaoka(2022)}]{okimoto2022csc}
--- (2022), \enquote{The Credit Spread Curve Distribution and Economic
  Fluctuations in Japan,} \textit{Journal of International Money and Finance},
  122, 102582.

\bibitem[{Ren and Shimotsu(2009)}]{ren2009ifs}
Ren, Y. and Shimotsu, K. (2009), \enquote{Improvement in Finite Sample
  Properties of the Hansen^^e2^^80^^93Jagannathan Distance Test,}
  \textit{Journal of Empirical Finance}, 16, 483--506.

\bibitem[{Roll(1977)}]{roll1977cap}
Roll, R. (1977), \enquote{A Critique of the Asset Pricing Theory's Tests Part
  I: On Past and Potential Testability of the Theory,} \textit{Journal of
  Financial Economics}, 4, 129--176.

\bibitem[{Ross(1976)}]{ross1976atc}
Ross, S.~A. (1976), \enquote{The Arbitrage Theory of Capital Asset Pricing,}
  \textit{Journal of Economic Theory}, 13, 341--360.

\bibitem[{Sargan and Bhargava(1983)}]{sargan1983mle}
Sargan, J.~D. and Bhargava, A. (1983), \enquote{Maximum Likelihood Estimation
  of Regression Models with First Order Moving Average Errors when the Root
  Lies on the Unit Circle,} \textit{Econometrica}, 51, 799--820.

\bibitem[{Schwarz(1978)}]{schwarz1978edm}
Schwarz, G. (1978), \enquote{Estimating the Dimension of a Model,}
  \textit{Annals of Statistics}, 6, 461--464.

\bibitem[{Shaikh(2019)}]{shaikh2019ore}
Shaikh, I. (2019), \enquote{On the Relationship between Economic Policy
  Uncertainty and the Implied Volatility Index,} \textit{Sustainability}, 11.

\bibitem[{Shanken(1992)}]{shanken1992ebp}
Shanken, J. (1992), \enquote{On the Estimation of Beta-Pricing Models,}
  \textit{Review of Financial Studies}, 5, 1--33.

\bibitem[{Sharpe(1964)}]{sharpe1964cap}
Sharpe, W.~F. (1964), \enquote{Capital Asset Prices: A Theory of Market
  Equilibrium under Conditions of Risk,} \textit{Journal of Finance}, 19,
  425--442.

\bibitem[{Silvennoinen and Thorp(2013)}]{silvennoinen2013fcc}
Silvennoinen, A. and Thorp, S. (2013), \enquote{Financialization, Crisis and
  Commodity Correlation Dynamics,} \textit{Journal of International Financial
  Markets, Institutions \& Money}, 24, 42--65.

\bibitem[{Thorbecke(2020)}]{thorbecke2020hcc}
Thorbecke, W. (2020), \enquote{How the Coronavirus Crisis Affected Japanese
  Industries: Evidence from the Stock Market,} RIETI Discussion Paper Series
  20-E-061.

\bibitem[{Thuraisamy et~al.(2013)Thuraisamy, Sharma, and
  Ahmed}]{thuraisamy2013rae}
Thuraisamy, K.~S., Sharma, S.~S., and Ahmed, H. J.~A. (2013), \enquote{The
  Relationship between Asian Equity and Commodity Futures Markets,}
  \textit{Journal of Asian Economics}, 28, 67--75.

\bibitem[{Tsuji(2007)}]{tsuji2007wmi}
Tsuji, C. (2007), \enquote{What Macro-Innovation Risks Really Are Priced in
  Japan ?} \textit{Applied Financial Economics}, 17, 1085--1099.

\bibitem[{Wen et~al.(2022)Wen, Shui, Chen, and Gong}]{wen2022mpu}
Wen, F., Shui, A., Chen, Y., and Gong, X. (2022), \enquote{Monetary Policy
  Uncertainty and Stock Returns in G7 and BRICS Countries: A
  Quantile-on-Quantile Approach,} \textit{International Review of Economics and
  Finance}, 78, 457--482.

\bibitem[{Yang et~al.(2019)Yang, Zhou, and Cheng}]{yang2019srg}
Yang, Z., Zhou, Y., and Cheng, X. (2019), \enquote{Systemic Risk in Global
  Volatility Spillover Networks: Evidence from Option-Implied Volatility
  Indices,} \textit{Journal of Futures Markets}, 40, 392--409.

\bibitem[{Zhang and Hamori(2021)}]{zhang2021com}
Zhang, W. and Hamori, S. (2021), \enquote{Crude Oil Market and Stock Markets
  during the COVID-19 Pandemic: Evidence from the US, Japan, and Germany,}
  \textit{International Review of Financial Analysis}, 74, 101702.

\end{thebibliography}
